# Adaptation and Optimization of Automatic Speech Recognition (ASR) for the Maritime Domain in the Field of VHF Communication

**Emin Çağatay Nakilcioğlu,** Fraunhofer CML, Hamburg/GERMANY emin.nakilcioglu@cml.fraunhofer.de
**Maximilian Reimann**, Fraunhofer CML, Hamburg/GERMANY, maximilian.reimann@cml.fraunhofer.de
**Ole John,** Fraunhofer CML, Hamburg/GERMANY ole.john@cml.fraunhofer.de

**Abstract**

*This paper introduces a multilingual automatic speech recognizer (ASR) for maritime radio communication that automatically converts received VHF radio signals into text. The challenges of maritime radio communication are described at first, and the deep learning architecture of marFM® consisting of audio processing techniques and machine learning algorithms is presented. Subsequently, maritime radio data of interest is analyzed and then used to evaluate the transcription performance of our ASR model for various maritime radio data.*

## 1. Introduction

Maritime communication is a critical aspect of global trade and transportation, enabling ships to communicate with one another as well as with shore-based facilities. VHF stands for Very High Frequency and is a widely used radio communication technology in the maritime domain, providing reliable voice communication. It is a type of electromagnetic radiation used mainly in the frequency range from 156.025 MHz to 162.025 MHz for maritime radio communications (*ITU, 2022)*. VHF radios are used for ship-to-ship, ship-to-shore (vice versa) and onboard communications (*Alagha and Løge, 2023)*. Due to its vast range, ease of use and technological robustness, VHF communication has been a useful instrument for decades and has become a part of the mandatory equipment on board according to SOLAS *(International Maritime Organization, 1974)*.

Despite its advantages, using VHF radio technology for communication poses several obstacles. A characteristic feature of VHF radio is the presence of background noise, which is further compounded by noise sources onboard ships, such as machinery or sea sounds. These can significantly impair speech intelligibility. Moreover, the signal quality is strongly affected by weather conditions, including fog, rain, and snow. Rain, in particular, can reduce radio wave strength and cause signal interruptions (*Meng et al., 2009)*. Weak connections between vessels and coastal stations, or interference from nearby sources, may also result in distortion or disruptions in radio calls.

In addition to these technological limitations, linguistic factors also influence the effectiveness of maritime radio communication. the internationality of the ship's crews introduces a wide range of language proficiencies, dialects, and accents that can complicate communication and reduce intelligibility (*John et al., 2017)*. Since English serves as the primary working language at sea, variation in accent strength and proficiency often affects clarity, especially when speakers are unfamiliar with diverse speech patterns. To reduce the problem of language barriers and to mitigate the occurrence of misunderstandings, the IMO introduced the "Standard Marine Communication Phrases" (SMCP) in 1977, which replaced the Standard Marine Navigational Vocabulary (SMNV). The SMCP is a framework containing phrases for routine situations and standard responses for emergency situations, which aims to reduce language barriers and avoid misunderstandings in maritime communication (*International Maritime Organization, 2001)*.

Despite its shortcomings, VHF radio is an indispensable tool for communication between ships and shore stations in the maritime domain and provides a fast, easy and reliable way to transmit information between different stakeholders. Automatic speech recognition technologies serve great potential to address the challenges in the field of VHF radio communication.

Automatic speech recognition (ASR) is an increasingly important component of human-computer interaction since it is a technology that enables machines to transcribe human speech into text. This technology has made considerable progress in recent years and has also found application in the end-consumer sector. Examples of areas where ASR is used today include voice control, voice input and automated customer support system, among others. ASR is based on machines' ability to recognize and interpret spoken language by calculating the most likely spoken sentence based on audio signals. This technology uses complex algorithms and mathematical models to capture the acoustic signals of speech and convert them into digital formats that can be processed (*Li, 2022*). However, applying ASR to the maritime domain presents unique and significant challenges. High levels of background noise from machinery, unpredictable signal quality, and a wide diversity of non-native accents are common. Furthermore, the lack of sizable, annotated maritime audio datasets makes it highly impractical to build top-performing models from scratch. Consequently, more robust approaches are required to leverage domain-specific knowledge.

Transfer learning provides an encouraging approach to overcome these domain adaptation challenges by repurposing previously trained models and fine-tuning them for domain-specific tasks. It has been used in various fields, including computer vision (*He et al., 2016)*, and natural language processing *(Peters et al., 2019)*. In the field of ASR, transfer learning has been used to improve the performance of models on low-resource languages (*Zhao and Zhang, 2022)*, and tasks (*Chorowski et al., 2019)*.

The Wav2Vec2 model, introduced by *Baevski et al. (2020),* is a state-of-the-art ASR model that utilizes self-supervised learning to learn speech representations from raw audio. The model uses a contrastive predictive coding (CPC) objective to learn contextualized representations of audio, which can then be used for downstream tasks. Fine-tuning the Wav2Vec2 model for downstream speech recognition tasks has shown significant improvements in its performance (*Baevski et al., 2020; Wang et al., 2021; Vaessen and Van Leeuwen, 2022; Zuluaga-Gomez et al., 2023)*.

Considering the international nature of maritime communication, ships and ports from different countries often communicate with each other, and there are a variety of languages used in maritime communication. A multilingual ASR system would enable efficient and accurate communication between different language speakers, improving safety and efficiency in the maritime domain.

In recent years, the development of multilingual ASR models has gained significant attention. With the addition of the recent advancements in deep learning, large-scale pre-trained models, such as the Cross-lingual Language Model Pre-training (XLM) and XLSR models, which have shown great potential for improving the performance of ASR systems, were introduced. The XLSR model, in particular, is a pre-trained model that has been trained on a large corpus of multilingual data and has shown to be effective in low-resource languages and domains. For example, in a novel study by *Conneau et al. (2020),* the XLSR models were pre-trained on a large corpus of speech data in hundreds of languages and achieved state-of-the-art performance on several benchmarks for cross-lingual speech recognition. Given the proven effectiveness of XLSR models in low-resource and cross-lingual scenarios, this approach stands out as a promising approach for tackling the challenges of maritime communication.

In this paper, we introduce marFM®, a multilingual ASR system for the maritime domain. Our approach is similar to transfer learning, as we use a pretrained XLSR model to initialize the weights of our model and then fine-tune it on our custom maritime audio database which consists of maritime audio recordings in English and German. By doing so, we aim to improve the accuracy and efficiency of ASR for maritime communication in multiple languages simultaneously and thus to facilitate safer and more efficient communication between vessels of different nationalities.

This paper is structured as follows. Section 2 provides the details about our implementation of supervised fine-tuning. The data we used and how we preprocessed it are also presented in this section. After a short introduction of the experiments in Section 3, we present and discuss the results in Section 4 followed by a conclusion in Section 5.

## 2. Methodology

This chapter lays out the underlying methodology applied in developing/training a multilingual ASR model for multilingual maritime communication. As shown in the Figure 1, our methodology consists of two main steps: data pre-processing and fine-tuning of a pretrained ASR-Model. In this section, the details of the fine-tuning process are presented, and our custom maritime database and the data preprocessing steps are discussed in detail.

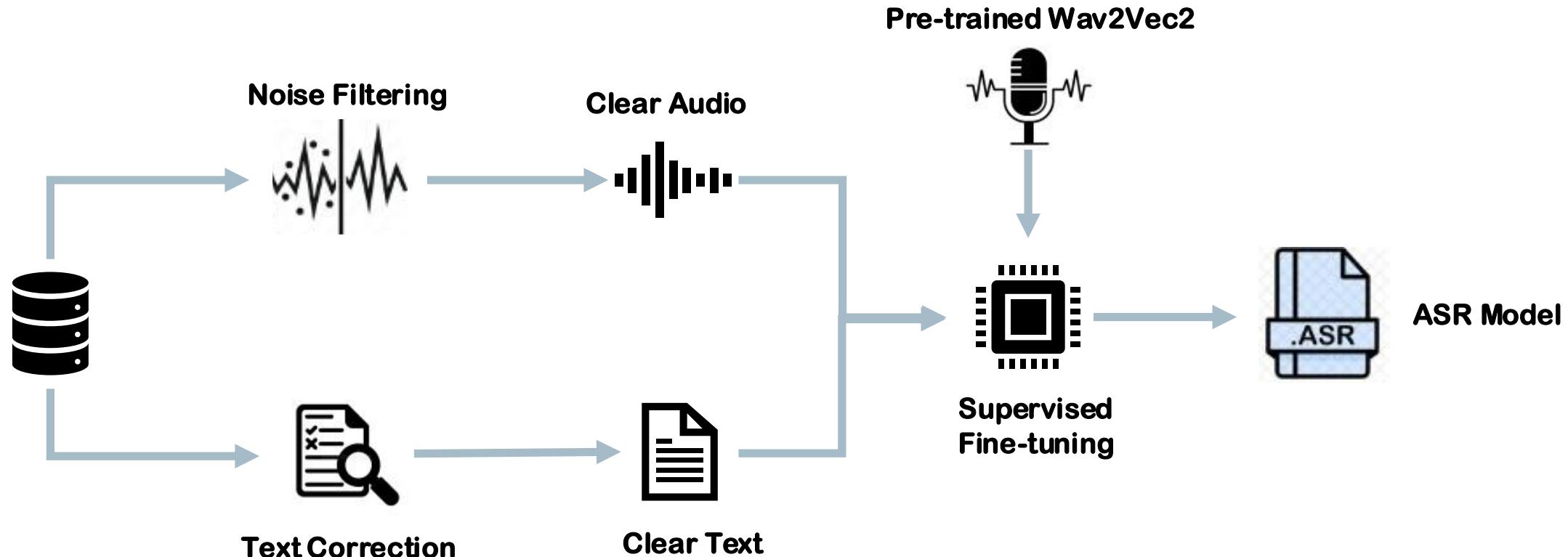

Figure 1. Preprocessing and fine-tuning of ASR-model.

### 2.1 Supervised Fine-Tuning

Our approach in this work is to use Wav2Vec2-XLSR-53 model pretrained on large-scale German data as our base model and to fine tune it on our custom maritime database. The pretrained model was acquired through Hugging Face Hub. Hugging Face is an open-source software library that provides easy-to-use interfaces for popular NLP models, including Wav2Vec2 and XLSR (*Wolf et al., 2020)*. It allows users to easily access and fine-tune these models on their own datasets. Hugging Face also provides pre-trained models that can be used as a starting point for fine-tuning. The pre-trained models are trained on large-scale datasets, which makes them suitable for transfer learning on smaller datasets. In our work, we utilized the Wav2Vec2-XLSR-53 German model from Hugging Face as our base model for fine-tuning on our maritime audio dataset.

The base architecture of the Wav2Vec2-XLSR-53 German model is based on the XLSR approach which was built on wav2vec 2.90 architecture and designed to learns cross-lingual speech representations by pretraining a single model from the raw waveform of speech in multiple languages (*Conneau et al., 2020).* As shown in Figure 2, the XLSR approach consists of a layer convolutional neural network (CNN) encoder followed by a transformer decoder. The CNN encoder extracts contextualized speech features from the input waveform whose embeddings are then act as targets for the training of the transformer decoder using contrastive learning. Similar to self-supervising training of wav2vec 2.0, XLSR only requires raw unlabelled speech audio in multiple languages. The Wav2Vec2-XLSR-53 German is pretrained on a large-scale German speech corpus and fine-tuned on the Common Voice German dataset, *Conneau et al. (2020)*.

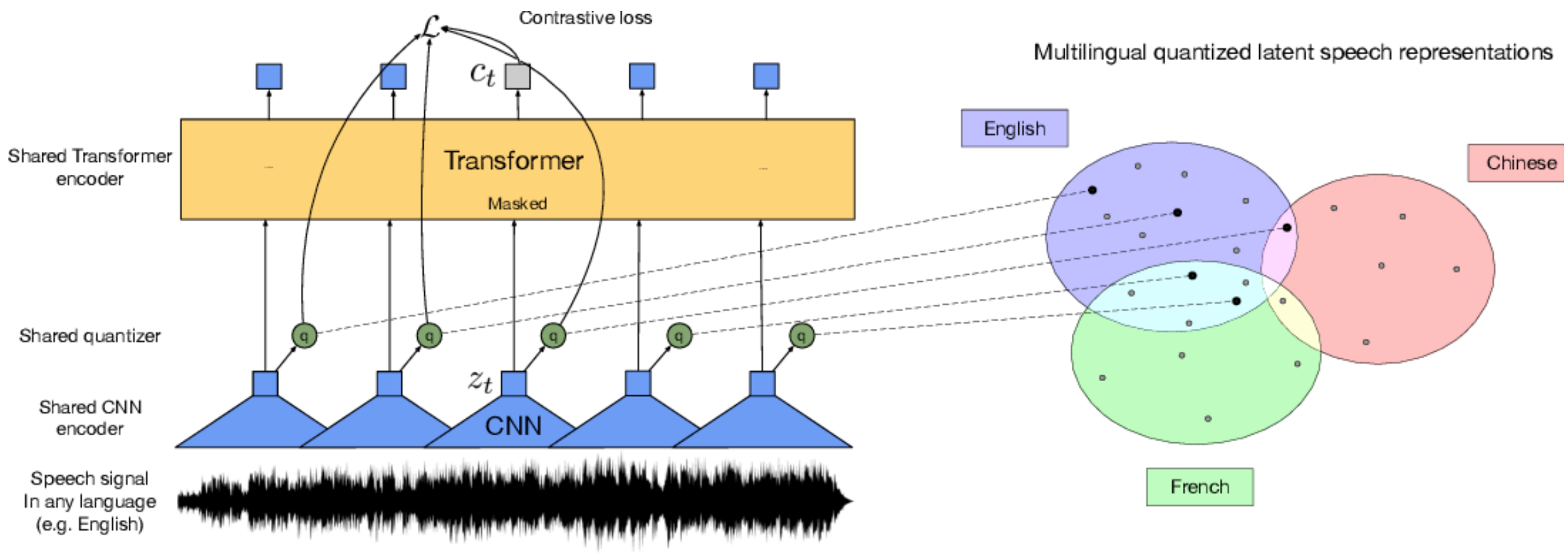

Figure 2. The XLSR approach on multilingual speech recognition, *Conneau et al. (2020)*

To fine-tune the Wav2Vec2-XLSR-53 model for the maritime domain, we used a custom maritime audio database that was collected by Fraunhofer CML. The database contains various types of maritime radio communication, such as distress calls, navigational messages, and weather reports as well as situational reports. The more detail regarding the database will be provided the following sub-section. For fine-tuning, we followed a transfer learning approach. We initialized the model with the pre-trained weights and fine-tuned it on our dataset using the Connectionist Temporal Classification (CTC) loss function. CTC loss function is a popular method for training ASR models, where the model learns to align the predicted output sequence with the ground truth by maximizing the log-likelihood of the correct transcription. This approach has been shown to be effective in various ASR tasks, including phone recognition (*Graves et al., 2006*), keyword spotting (*Michaely et al., 2017*), and speaker diarization (*Wang et al., 2021*). In our work, we use CTC loss to train our multilingual ASR model to predict character sequences from audio waveforms. We use the CTC loss implementation provided by the PyTorch framework (*Paszke et al., 2019*). During the fine-tuning process, we used a learning rate of 3e-5, a batch size of 8. We also used early stopping to prevent overfitting and reduce training time. The training for fine-tuning on a single GPU (Nvidia RTX A6000) took approximately 34 hours in total.

## 2.2 Database and Preprocessing

For this project, Fraunhofer CML have created our own custom maritime database by gathering maritime audio recordings. This database consists of about 62-hour long audio recordings of real VHF radio conversations in English and German together with their respective transcriptions.

Since the pretrained XLRS model was trained on data with a sampling rate of 16 kHz, the fine-tuning data needed to be converted into the same format so that the convolutional filters could operate on the same timescale. To this end, every audio recording in the database was resampled to 16 kHz. Preprocessing was done using librosa library in Python, *McFee et al. (2015)*.

As mentioned in Section 1, another main challenge regarding VHF-Radio recordings is the heavy background noise that is ever present regardless of the type of VHF-Radio hardware. Depending on the range of the VHF-Radio receivers, the hardware quality and the distance between the communicators, prepotency of the background differs in the recordings. To address the background noise with different characteristics, we applied non-stationary noise gating to the raw audio data. The "noisereduce" library in Python was used for noise gating/reduction (*Sainburg, 2019)*. The library offers two different noise reduction algorithms: stationary and non-stationary. Main difference between the algorithms is that non-stationary noise reduction allows the noise gate to change over time by using a sliding window where noise statistics are recomputed every time the window is moved to adjacent part of the audio clip. The working schematics of stationary and non-stationary noise reduction is shown in Figure 3. Following steps are applied during the noise reduction process.

- First, a spectrogram is calculated over the signal
- A time-smoothed version of the spectrogram is computed using an IIR filter applied forward and backward on each frequency channel,
- A mask is computed based on that time-smoothed spectrogram
- The mask is smoothed with a filter over frequency and time
- The mask is applied to the spectrogram of the signal, and is inverted

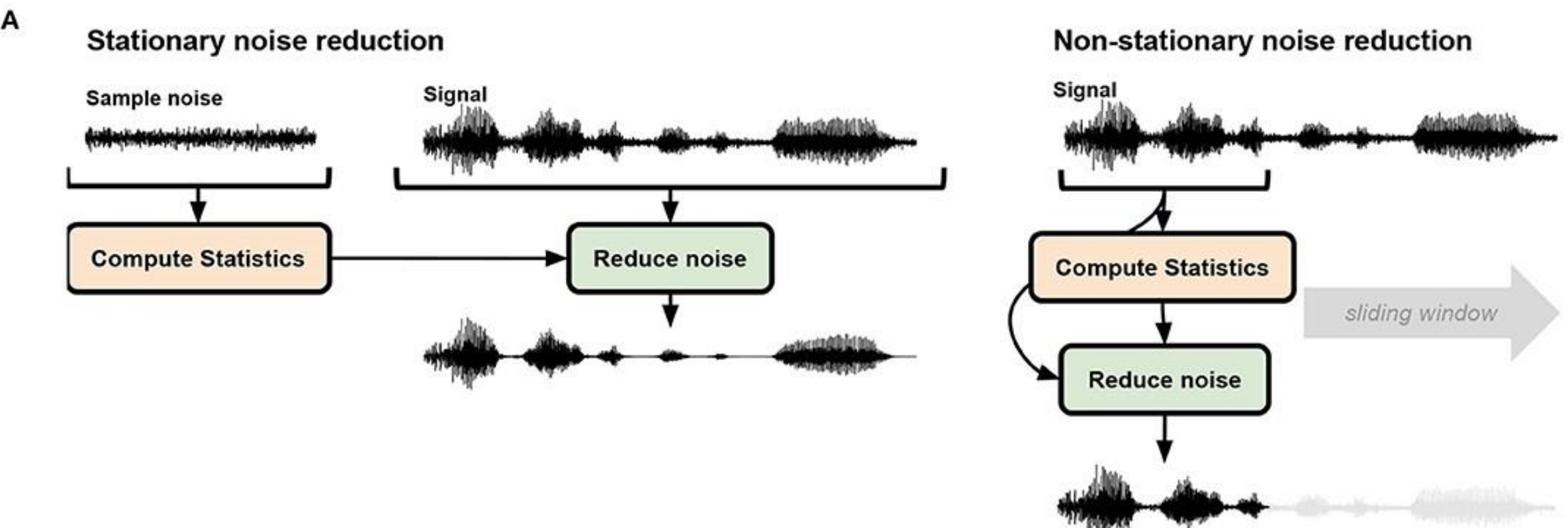

Figure 3. Stationary vs. non-stationary noise reduction (*Sainburg and Gentner, 2021).*

The other preprocessing steps, text correction and clearing, were also carried out simultaneously. The objective behind text correction and clearing was removing all characters that don't contribute to the meaning of a word and cannot be properly represented by an acoustic sound and thus, normalizing the text. A quick investigation revealed that our transcriptions contained some punctuation marks which do not correspond to any characteristic sound unit, i.e., phoneme. For example, the letter "d" has its own phonetic sounds whereas the punctuation mark "-" cannot be represented as a phoneme. To finalize the text clearing stage, we have consulted with a native German speaker to find out whether any further simplification and clearing of the transcriptions was possible. We were informed that the certain vowels that were to be written with an umlaut such as ä, ö and ü were written according to the English alphabets. Thus, as the final step of the preprocessing, we rewrote these vowels according to the German alphabet.

## 3. Experiments

For the experimental setup, we have split the dataset into, training and test datasets with the ratio of 90:10, which correspondents to ca. 56 h of audio for training and 6 h of audio for testing. Training dataset was further divided into two splits: train and validation sets. Ratio of 80:20 was applied to these splits which adds up to ca. 45 h of audio for training and 11 h of audio for validation. In summary, 6 hours of maritime recordings consisting of real VHF radio calls both in English and German were used to evaluate the model performance.

Three different models were considered for the comparative performance analysis. Our base model, Wav2Vec2-XLSR-53 German, was involved in the experiments in order to see the full effect of the fine tuning on the model's performance. We have also evaluated the performance of the Wav2Vec2-XLSR-53 German with a language model. As shown in *Udagawa et al. (2022)*, *Vaibhav and Padmapriya (2022)* and on model evaluation pages of Hugging Face Hub, attaching a language model to a Wav2Vec2 model has a positive effect on the decoding performance of the ASR models. Thus, we have also included Wav2Vec2-XLSR-53 German with a language model trained on English and German phrases.

In late 2022, a general-purpose speech recognition model called Whisper were introduced by OpenAI Research Lab (*Radford et al., 2023)*. It is trained on a large dataset of diverse audio, 680,000 hours of multilingual and multitask supervised data collected from the web and is also a multi-task model that can perform multilingual speech recognition as well as speech translation and language identification. The Whisper's zero-shot performance were measured across many diverse datasets, and it showed is much more robustness and made fewer errors than the other state-of-the-art models. We have concluded that the involvement of a Whisper-based model to our experimental setup would be worthwhile addition and thus, in the evaluation phase, we have also considered the transcription performance of the largest available Whisper model, Whisper large-v2 model, which is biggest model in size and trained for 2.5x more epochs with added regularization for improved performance compared Whisper large model.

Speech recognition research typically evaluates and compares systems based on the word error rate (WER) metric (*Woodard and Nelson, 1982*). To calculate WER, the reference and recognized transcript are aligned. It is accomplished by minimizing the Levenshstein distance (or edit distance) between the two texts. Following this, you can count the number of substitutions (S), insertions (I), and deletion (D) errors. In its simply form, the WER is the proportion of the total number of errors over the number of words in the reference. In a transcription output where all the words were correct, the WER would be zero which we are aiming to drive towards in terms of the marFM®'s performance. The WER calculated using the following equation

$$WER = \frac{S + D + I}{N}$$

Equation 1: Word error rate formula.

where N being the total number of symbols in the reference word, D representing the number of deleted symbols in the hypothesis with respect to the reference word, S being the number of changed symbols, and I being the number of additional symbols.

## 4. Results

In this study, we compared the multilingual speech recognition performance between four different models including marFM® on the test split of our custom maritime database which contains 6 hours of real VHF radio calls. Wav2Vec2-XLSR-53 German model performance was evaluated both with and without an additional language model (LM). Other models performed the multilingual transcription task without the attachment of any external language model.

The results of the multilingual audio transcription task are presented in Table I. Each model used a shared capacity of a single GPU (Nvidia RTX A6000) while generating their predictions, i.e., transcriptions. The total inference time of the models added up to approximately 19 minutes.

As expected, the positive influence of attaching a language model to the base model, Wav2Vec2-XLSR-53 German, was apparent on the model's transcription performance. With the help of the language model, its transcription accuracy has increased by approximately 4%. The base model without any language model attached ended up performing the worst among the models. Considering that the dataset the model was trained on consisted of mostly daily speech recordings, the relatively poor performance by the base model was expected. However, though it was trained on a data with similar characteristics, Whisper large-v2 model has outperformed both XLSR models. We observed close to 7% improvement in accuracy when compared to the base model and about 3% compared to the XLSR model with the language model extension. Although both architectures, Wav2Vec2 and Whisper, were based on transformers and encoder-decoder style approach, the Whisper implementation has demonstrated a certain level of superiority regarding the transcription accuracy on the multilingual VHF calls.

Table I: Results of multilingual speech recognition performance among different ASR models

| Model | Word Error Rate (WER%) |
| --- | --- |
| Wav2Vec2-XLSR-53 DE | 44.36 |
| Wav2Vec2-XLSR-53 DE w/ ML | 40.58 |
| Whisper Large V2 | 37.62 |
| **marFM®** | **31.59** |

Despite the higher accuracy output of the Whisper large-v2 model, marFM® has produce the most accurate transcriptions among the models. With 31.5 % of WER, marFM® has shown a significant improvement in comparison to its closest contender which is the Whisper large-v2 model with 37.62 % of WER. As hypothesized, fine-tuning process has elevated the model's transcription performance with an approximately 13% boost in WER in comparison to the transcription accuracy of the base model, Wav2Vec2-XLSR-53 German.

## 5. Conclusion and Future Work

In this study, we introduced marFM® a multilingual automatic speech recognizer for maritime communications. We also showcased its superior performance on the multilingual transcription task for VHF radio calls in comparison with the other open-source and readily available ASR models.

One of the main challenges for developing an ASR system for maritime domain was collecting the real maritime data with its corresponding transcriptions for our fine-tuning process. Throughout the span of multiple projects, we were able to gather a total of 62 hours of maritime recordings. Thanks to the state-of-the-art ASR architectures such as Wav2Vec2 and XLSR, we were able to develop marFM® and to provide an ASR system transcribing the maritime calls with higher accuracy compared to the other open-source, state-of-the-art ASR systems available.

As the experimental results have demonstrated, the addition of a language model can help Wav2Vec2-based ASR models transcribe with more accuracy. Thus, the performance of marFM® is expected to increase as we attach a language model trained on a language dataset expanded with maritime-related corpus. For this end, we have initialized a development process where we started training our own language model with the maritime expansion.

Among the non-finetuned models, Whisper large-v2 model has shown the best transcription performance. Thus, we are of the opinion that it would be worthwhile research effort to develop Whisper-based maritime ASR in order to evaluate whether we can achieve higher accuracy levels with such implementation for maritime communication. For this end, we have also initialized a side research project where we explore the capabilities of Whisper in the maritime domain further.